\documentclass[12pt,preprint]{aastex}

\usepackage{color}
\usepackage{ulem}

\slugcomment{\today}

\shorttitle{Extensive [CI] Mapping toward the Orion-A Molecular Cloud}
\shortauthors{Shimajiri et al.}

\begin{document}


\title{Extensive [CI] Mapping toward the Orion-A Giant Molecular Cloud}


\author{Yoshito Shimajiri,\altaffilmark{1,2,3} Takeshi Sakai,\altaffilmark{4,5} Takashi Tsukagoshi,\altaffilmark{6} Yoshimi Kitamura,\altaffilmark{7} Munetake Momose,\altaffilmark{6} Masao Saito,\altaffilmark{2,8} Tai Oshima,\altaffilmark{1,2} Kotaro Kohno,\altaffilmark{4,9} and Ryohei Kawabe\altaffilmark{1,2,8}}

\email{Yoshito.Shimajiri@cea.fr}



\altaffiltext{1}{Nobeyama Radio Observatory, 462-2 Nobeyama Minamimaki, Minamisaku District, Nagano Prefecture 384-1305, Japan}
\altaffiltext{2}{National Astronomical Observatory of Japan, 2-21-1 Osawa Mitaka, Tokyo 181-0015, Japan}
\altaffiltext{3}{Laboratoire AIM, CEA/DSM-CNRS-Universit$\acute{e}$ Paris Diderot, IRFU/Service d'Astrophysique, CEA Saclay, F-91191 Gif-sur-Yvette, France}
\altaffiltext{4}{Institute of Astronomy, The University of Tokyo, 2-21-1 Osawa, Mitaka, Tokyo 181-0015, Japan}
\altaffiltext{5}{Graduate School of Informatics and Engineering, The University of Electro-Communications, Chofu, Tokyo 182-8585, Japan}
\altaffiltext{6}{Ibaraki University, 2-1-1 Bunkyo Mito, Ibaraki Prefecture 310-8512, Japan}
\altaffiltext{7}{Institute of Space and Astronautical Science, Japan Aerospace Exploration Agency, 3-1-1 Yoshinodai, Chuo-ku, Sagamihara 252-5210, Japan}
\altaffiltext{8}{Joint ALMA Observatory, Alonso de Cordova 3107 Vitacura, Santiago 763 0355, Chile}
\altaffiltext{9}{Research Center for the Early Universe, The University of Tokyo, 7-3-1 Hongo, Bunkyo, Tokyo 113-0033, Japan}


\begin{abstract}
We have carried out wide-field (0.17 degree$^2$) and high-angular resolution (21.3$\arcsec$ $\sim$ 0.04 pc) observations in [CI] line toward the Orion-A giant molecular cloud with the Atacama Submillimeter Telescope Experiment (ASTE) 10 m telescope in the On-The-Fly (OTF) mode. 
Overall features of the [CI] emission are similar to those of the $^{12}$CO ($J$=1--0) emission in \citet{Shimajiri11}; the total intensity ratio of the [CI] to CO emission ranges from 0.05 to 0.2.
The optical depth of the [CI] emission is found to be 0.1 -- 0.75, suggesting optically thin emission. The column density of the [CI] emission is estimated to be (1.0 -- 19) $\times$ 10$^{17}$ cm$^{-2}$.
These results are consistent with the results of the previous [CI] observations with a low-angular resolution of  2.2$\arcmin$ (e.g.  \citet{Ikeda99}). 
In the nearly edge-on PDRs and their candidates of the Orion Bar, DLSF, M 43 Shell, and Region D, the distributions of the [CI] emission coincide with those of the $^{12}$CO emission, inconsistent with the prediction by the plane-parallel PDR model \citep{Hollenbach99}. 
In addition, the [CI] distribution in the Orion A cloud is found to be more similar to those of the $^{13}$CO ($J$=1--0), C$^{18}$O ($J$=1--0), and H$^{13}$CO$^+$ ($J$=1--0) lines than that of the $^{12}$CO ($J$=1--0) line, suggesting that the [CI] emission is not limited to the cloud surface, but is tracing the dense, inner parts of the cloud.
\end{abstract}


\keywords{ISM: clouds \ ISM: individual (Orion-A giant molecular cloud) \ ISM: atomic carbon}



\section{Introduction}
Far ultraviolet (FUV: 6 eV $<$ $h$$\nu$ $<$ 13.6 eV) radiation emitted from massive stars influences the structure, chemistry, thermal balance, and evolution of the neutral interstellar medium of galaxies \citep{Hollenbach97}. 
Hence, studies of the influence of FUV are crucial to understanding the process of star formation.
Regions where FUV photons dominate the energy balance or chemistry of the gas are termed Photon-Dominated Regions (PDRs). 
The plane-parallel models of the PDRs predict layers of species such as H/H$_2$, C$^+/$C/CO, and others \citep{Hollenbach99}. 
Since the energy required to ionize carbon (11.3 eV) is close to that required to photodissociate CO (11.09 eV), 
the [CI] emission should exist over a fairly narrow range of column densities, sandwiched between the C$^+$ layer and the CO layer \citep{Tielens85a,Tielens85b}. 
To observationally detect the thin C layer, one should observe the PDR that is influenced by the FUV emission from an exciting star in a nearly edge-on configuration \citep{Gorti02}. There are, however, relatively few observations toward the PDRs irradiated in nearly edge-on configuration in the [CI] line, owing to the difficulty of the observations with ground-based telescopes. 

The Orion-A giant molecular cloud (Orion-A GMC), with an extent of $\sim$10 degrees (corresponding to $\sim$72 pc) in the southern part of the Orion constellation, is the nearest ($d$= 400 pc; \cite{Menten07,Sandstrom07,Hirota08}) GMC and one of the best studied star-forming regions \citep{Tatematsu99,Johnstone99,Shimajiri08,Shimajiri09, Takahashi08,Davis09}. In the northern part of the Orion-A GMC, 
there exist the three HII regions, M 42, M 43, and NGC 1977 \citep{Goudis82}. 
\citet{Shimajiri11} have already identified six PDR candidates in the northern part of the Orion-A GMC from the comparison among the distributions of the AzTEC 1.1-mm, Nobeyama 45 m $^{12}$CO ($J$=1--0), and Midcourse Space Experiment (MSX) 8 $\mu$m emissions. 
Since the stratification among the distributions of the 8 $\mu$m (Polycyclic Aromatic Hydrocarbons, PAHs), 1.1-mm dust continuum, and $^{12}$CO emissions can be recognized, the PDR candidates are likely influenced by the far-ultraviolet (FUV) emission from the Trapezium star cluster and from NU Ori in nearly edge-on configuration.
Thus, the Orion-A GMC is one of the most suitable targets to investigate PDRs.
Previous studies carried out [CI] observations toward the Orion-A GMC \citep{White95,Ikeda99,Plume00}. 
\citet{Ikeda99} observed the [CI] line toward an approximately nine square degree area of the Orion-A GMC with a low-angular resolution of  2.2$\arcmin$ and a grid spacing of 3$\arcmin$ (corresponding to a spatial dynamic range $A_{\rm map}$/$A_{\rm beam}$\footnote[+1]{We refer to the telescope beam size as $A_{\rm beam}$.} = 2130). \citet{White95} observed the [CI] line with a high-angular resolution of  9.8$\arcsec$ and a grid of 10$\arcsec$ toward a small ($\sim$3$\arcmin$ $\times$ 1.5$\arcmin$) area of the OMC-1 region (corresponding to $A_{\rm map}$/$A_{\rm beam}$$^1$ = 50).  
[CI] observations toward the PDRs in the Orion-A GMC with both a high-angular resolution of $\sim$10$\arcsec$ and a large dynamic range of $\sim$1000, have not yet been performed, however.

With the advent of low-noise receivers at a frequency of $\sim$500 GHz, wide-field mapping observations of the [CI] emission with high angular resolution are now feasible. The ALMA Band 8 QM receiver has been installed on the Atacama Submillimeter Telescope Experiment (ASTE) telescope from October 2010 \citep{Satou08}. This receiver is a dual-polarization side-band separating (2SB) mixer receiver operating at 400 -- 500 GHz, developed as the ALMA Band 8 QM.  This equipment enables us to observe the [CI] line in the On-The-Fly (OTF) mode. 
In this paper, we present the first Nyquist-sampled [CI] image of the Orion-A GMC covering a wide field (0.17 degree$^2$) with an effective angular resolution of 21.3$\arcsec$ giving a spatial dynamic range $A_{\rm map}$/$A_{\rm beam}$$^1$ = 3680, including four PDR candidates in nearly edge-on configurations.

\section{Observations}

The [CI] ($^3$P$_1$-- $^3$P$_0$; 492.1607 GHz) data in the Orion-A GMC have been taken with the ASTE 10 m telescope \citep{Ezawa04,Kohno04} located at the Pampa la Bola at an altitude of 4800 m. 
The observations were carried out remotely from the ASTE operation room at Mitaka, Tokyo in October and November 2010, using the network observation system N-COMOS3 developed by NAOJ \citep{Kamazaki05}.
The typical system noise temperature with the 500 GHz SIS ALMA Band 8 QM receiver in SSB mode was 500 -- 2500 K, and the atmospheric opacity at 220 GHz was 0.01 -- 0.07 during our observations. The temperature scale was determined by the standard chopper-wheel method, which provides us with the antenna temperature corrected for the atmosphere attenuation. As a back end, we used four sets of a 1024 channel auto-correlator, which gives a frequency resolution of 0.5 MHz, corresponding to 0.3 km s$^{-1}$ at the [CI] frequency. 
The OTF mapping technique was employed for the [CI] mapping over a 0.17 degree$^2$ region. The telescope pointing was checked every two hours, and the pointing errors were measured to be $\sim$2$\arcsec$ during the observations. 
The main-beam efficiency of the ASTE telescope was measured to be 49.2 \% at 492 GHz by observing Jupiter.
After subtracting linear baselines, the OTF data were spatially smoothed by a spheroidal function$\footnote[+2]{\citet{Schwab84} and  \citet{Sawada08} described the details of the spheroidal function. We applied the parameters, $m$ = 6 and $\alpha$ = 1, which define the shape of the function.}$
with a FWHM (full width at half maximum) of 10$\arcsec$ to make cube data. 
Since the telescope beam pattern at 492 GHz was approximately described by a Gaussian function with a FWHM of 13.8$\arcsec$, the final resolution of the cube data becomes 21.3$\arcsec$ in FWHM.
The scanning effect was minimized by combining scan along both the R.A. and decl. directions. The typical rms noise level in the final cube data is 0.2 K in $T_{\rm A}$$^{*}$ with a velocity resolution of 0.6 km s$^{-1}$.

\section{Results}
\subsection{Distribution of the [CI] ($^3$P$_1$-- $^3$P$_0$) Emission }

Figures \ref{CIMOM0} (a) and (b) show total integrated intensity maps of the  [CI] ($^3$P$_1$-- $^3$P$_0$) and $^{12}$CO ($J$=1--0), respectively, over a velocity range 3.75 -- 16.25 km s$^{-1}$ in $V_{\rm LSR}$ toward the northern part of the Orion-A GMC: the $^{12}$CO map is from \citet{Shimajiri11}.
The overall distribution of the [CI] emission is found to be similar to that of the $^{12}$CO emission. 
The brightest peak in the [CI] map is located near Ori-KL; the peak positions of the $^{12}$CO and [CI] emissions are (RA, Dec) = (5$^{\rm h}$35$^{\rm m}$14$^{\rm s}$.6, -5$^{\circ}$22$\arcmin$21.2$\arcsec$) and (5$^{\rm h}$35$^{\rm m}$13$^{\rm s}$.7, -5$^{\circ}$21$\arcmin$52.4$\arcsec$), respectively (see also Figure \ref{CIMOM0} (c)).
The positional offset of 32$\arcsec$ between the $^{12}$CO and [CI] peaks is significantly larger than the positional uncertainties in the [CI] and $^{12}$CO observations, and is consistent with the previous study by \citet{White95}. 
Toward the south-east of Ori-KL, there exists a bar-like feature with a length of $\sim$0.6 pc. This feature corresponds to the Orion Bar, clearly recognizable in the $^{12}$CO peak map in Figure \ref{CIMOM0} (c). 
The second-strongest [CI] peak is seen on the western side of the Orion Bar and has a position of (RA, Dec)=(5$^{\rm h}$35$^{\rm m}$13$^{\rm s}$.8, -5$^{\circ}$26$\arcmin$44.0$\arcsec$), where no bright peak in the $^{12}$CO emission is seen.
Hereafter, we refer to the peak as CI PEAK.
In the eastern part of the [CI] map, there exists a filamentary structure along the north-south direction, which includes the Dark Lane South Filament (DLSF) identified as a PDR by \citet{Rodriguez01}.
The distribution of the [CI] emission in the filament is quite similar to that of the $^{12}$CO emission. 
On the western side of the DLSF, one can see a weak [CI] peak associated with Region D (see Figure \ref{8um} (d)), which is identified as a PDR candidate irradiated by $\theta^1$Ori C \citep{Shimajiri11}. 
To the south of Ori-KL and CI PEAK, there is a bright area with a peak intensity of 10 K.
This area is called OMC-4, which is a part of the integral-shaped filament \citep{Johnstone99}. 
In the total intensity map of the $^{12}$CO emission, a shell-like structure on the west of NU Ori is seen. 
This shell is known as a PDR irradiated by NU Ori \citep{Rodriguez01, Shimajiri11}.
The [CI] emission can be recognized toward the $^{12}$CO shell-like structure on the west of NU Ori, as shown in Figure \ref{CIMOM0} (a). 

Figure \ref{CIMOM0} (d) shows a total intensity ratio map between the [CI] and $^{12}$CO emissions, $I_{\rm [CI]}$/$I_{\rm CO}$. 
There is a spatial variation in the total intensity ratio. 
In the western part of the observed area, $I_{\rm [CI]}$/$I_{\rm CO}$ ranges from 0.1 to 0.2 with the latter value being the highest in the map.
In the northern parts of Ori-KL and DLSF, $I_{\rm [CI]}$/$I_{\rm CO}$ ranges from 0.1 to 0.18 with the latter value being relatively high in the map.
In the eastern parts of Orion Bar and M 43 Shell, $I_{\rm [CI]}$/$I_{\rm CO}$ has lower values of 0.06 -- 0.1.

\subsection{Velocity Structure of the [CI] ($^3$P$_1$-- $^3$P$_0$) Emission}

Here, we compare the spectra of the [CI] and $^{12}$CO ($J$=1--0) emissions to investigate the difference 
between the velocity structures of the atomic and molecular gases.
Figures \ref{spectra} (a)--(f) show the spectra of the [CI] and $^{12}$CO emissions at the six positions of M 43 Shell, Ori-KL, CI PEAK, OMC-4, Region D, and DLSF, respectively, shown in Figure \ref{CIMOM0} (c).
At each position, the overall velocity structure of the [CI] emission is similar to that of the $^{12}$CO ($J$=1--0) emission. 
Figure \ref{spectra} (b) shows the [CI] and $^{12}$CO spectra at Ori-KL. 
The $^{12}$CO emission has a very wide velocity width, indicating the presence of the molecular outflow from Ori-KL \citep{Allen93, Furuya09}. 
The velocity width of the [CI] emission also becomes the widest among the six [CI] spectra. 
The peak velocities of the two lines, however, are slightly different from each other: 9.4 ${\pm \ 0.4}$ and 8.1 ${\pm \ 0.6}$ km s$^{-1}$ for the $^{12}$CO and [CI] lines, respectively. 
In OMC-4, the peak intensity ratio of the [CI] to $^{12}$CO emissions reaches its maximum value of 0.21 (see Figure  \ref{spectra} (d)).
This is the highest ratio among six positions. 
In Region D, the $^{12}$CO emission seems to have two velocity components with peak velocities of 8.2 ${\pm \ 0.4}$ and 10.2 ${\pm \ 0.4}$ km s$^{-1}$, as shown in Figure \ref{spectra} (e). The [CI] emission also has two components at velocities  7.5 ${\pm \ 0.6}$ and 9.9  ${\pm \ 0.6}$ km s$^{-1}$. 
The $^{12}$CO emission at the DLSF in Figure \ref{spectra} (f) seems to have two velocity components: the peak velocities of the two components are 7.8 ${\pm \ 0.4}$ and 10.6 ${\pm \ 0.4}$ km s$^{-1}$. 
The [CI] emission also has two velocity components with 8.6 ${\pm \ 0.6}$ and 11.1${\pm \ 0.6}$ km s$^{-1}$. 
The 11.1 km s$^{-1}$ component is detected at a signal-to-noise ratio of 3.5.

\subsection{Optical Depth of the [CI] emission and [CI] Column Density}

We estimated the [CI] column density from the [CI] intensity under the assumption of local thermodynamic equilibrium (LTE) using equations (2) - (5) in \citet{Oka01}.
Here, we considered the peak intensity in $T_{\rm MB}$ of the $^{12}$CO ($J$=1--0) emission as the [CI] excitation temperature, assuming that the $^{12}$CO emission is optically thick. 
Figures \ref{CIMOM0} (e) and (f) show the distributions; the optical depth and the column density maps range from 0.1 -- 0.8 and (1.0 -- 19) $\times$ 10$^{17}$ cm$^{-2}$, respectively. 
Table \ref{properties} shows the optical depth and column density of [CI] toward the six positions shown in Figure \ref{CIMOM0} (c). The [CI] column density, $N_{\rm CI}$, is (2.3 -- 16) $\times$ 10$^{17}$ cm$^{-2}$. The optical depth and column density at Ori-KL are consistent with those in the previous [CI] study by \citet{Ikeda99}. Since the [CI] emission is optically thin all over the observed region, the [CI] column density becomes roughly proportional to the [CI] total intensity, as shown in Figure \ref{CIMOM0} (f). 
On the western side of the map, however, the [CI] column density becomes relatively high compared with the [CI] total intensity, because the optical depth of the [CI] emission is relatively high (0.3 -- 0.8) there.
This high-column density area can be seen as a shell-like structure with a size of $\sim$2 pc in the $^{12}$CO ($J$=1--0) channel maps of Figure 6 by \citet{Shimajiri11}. 
Since the dense (10$^{4-5}$ cm$^{-3}$) gas traced by the C$^{18}$O ($J$=1--0) and H$^{13}$CO$^{+}$ ($J$=1--0) emissions is not detected in the shell-like structure, the density of the shell is likely as low as 10$^3$ cm$^{-3}$.

\section{Discussion --What does the [CI] emission trace?--}

The [CI] emission line is thought to be one of the best tools to probe the structures of PDRs. 
In our observed area, there are three PDRs -- Orion Bar, M 43 Shell, and DLSF, and the one candidate -- Region D. Orion Bar, DLSF, and Region D are irradiated by $\theta^1$Ori C and M 43 Shell is irradiated by NU Ori. 
Thus, our PDR targets are most likely to be edge-on and actually the $^{12}$CO and 8 $\mu$m emission layers are clearly separated from each other, as shown in Figure 3. The 8 $\mu$m emission mostly arises from PAHs, and has been observed at the boundary of the H II regions \citep{Zavagno06}.
According to the plane-parallel model for a PDR \citep{Hollenbach99, Gorti02}, the [CI] emission layer should be clearly seen on the $^{12}$CO cloud surface that faces the exciting star in an edge-on configuration.
Although our spatial resolution was as high as 0.04 pc, no stratification can be seen between the [CI] and $^{12}$CO ($J$=1--0) emissions in Figure \ref{8um}. 
On the contrary, the distribution of the [CI] emission coincides with those of the $^{12}$CO emission, as shown in \citet{Ikeda99}. Intensity profiles of the [CI], $^{12}$CO, and MSX 8 $\mu$m emissions between the exciting stars ($\theta$$^{1}$Ori C or NU Ori) and PDRs (Orion Bar, M 43 Shell, DLSF Region D) show the coexistence of the [CI] and $^{12}$CO gas and   
the non-coexistence of PAHs and $^{12}$CO gas.  A range of separation between the peak position of $^{12}$CO and 8 $\mu$m emissions among four regions is 0.05 ${\pm \ 0.02}$ pc, which is larger than the beam size of  0.04 pc in the $^{12}$CO map. 
The coexistence of the [CI] and CO gas has been reported also in the AFGL 333 cloud \citep{Sakai06}. 
The coexistence disagrees with the prediction of the plane-parallel homogeneous model, and requires a clumpy PDR model.  
The UV radiation can penetrate much deeper in a clumpy cloud \citep{Spaans96,Kramer08}, predicting that the [CI] emission can be distributed over the whole cloud.

\begin{figure}
\epsscale{1.0}
\plotone{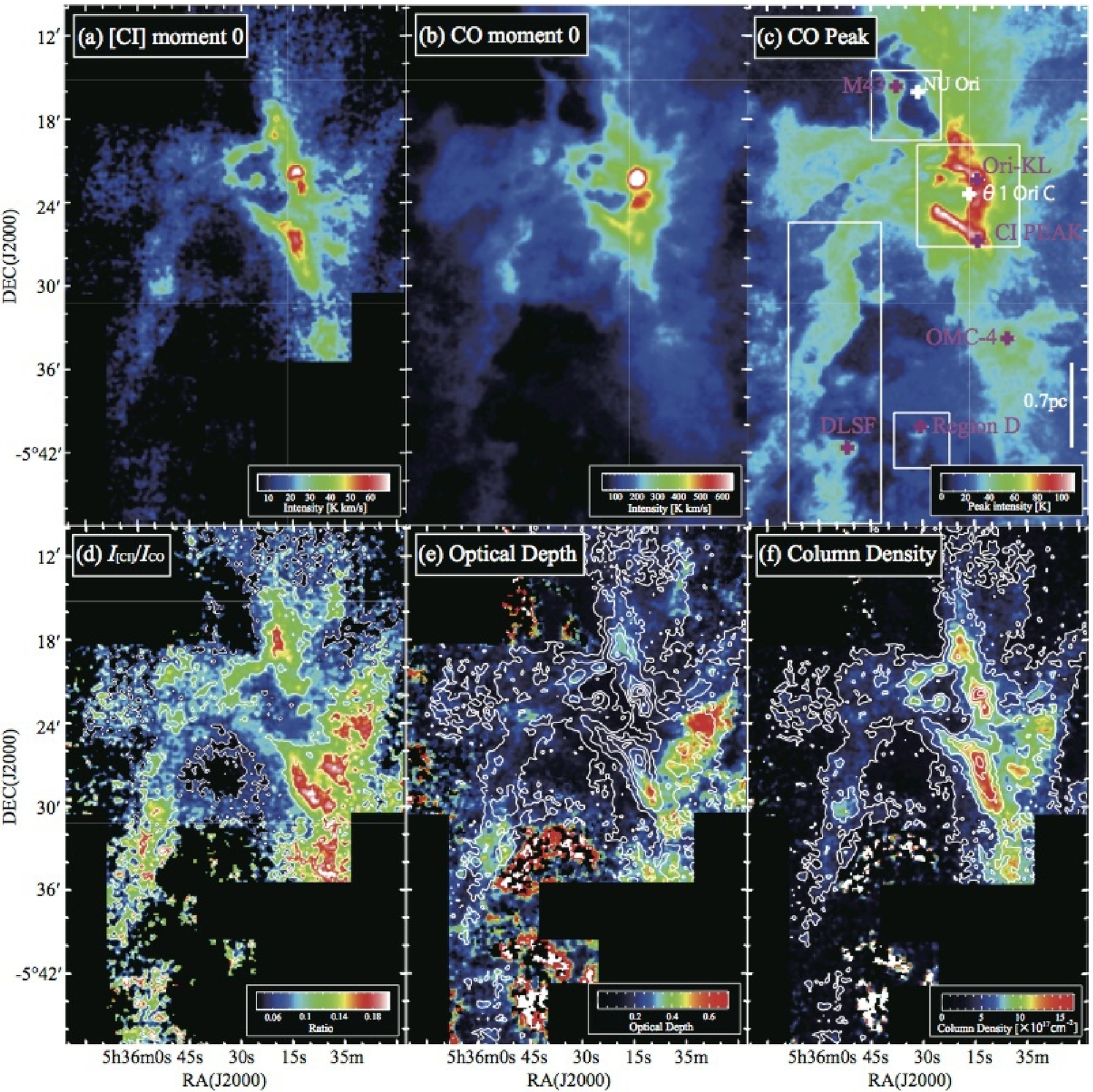}
\caption{
(a) [CI] and (b) $^{12}$CO ($J$=1--0) total integrated intensity maps over a $V_{\rm LSR}$ range of 3.75 -- 16.25 km s$^{-1}$ in units of K km s$^{-1}$, (c) $^{12}$CO peak intensity map in units of K, (d) $I_{\rm [CI]}$/$I_{\rm CO}$ map of the intensity ratio of the [CI] to $^{12}$CO emission, (e) optical depth map of the [CI] emission, and (f) [CI] column density map in units of 10$^{17}$ cm$^{-2}$. 
In the panel (c), the magenta plus symbols show positions for the spectra in Figure \ref{spectra} at Ori-KL, CI PEAK, DLSF, OMC-4, M 43 Shell, and Region D and the white plus ones the positions of the PDR exciting stars, $\theta^{1}$Ori C and NU Ori.
The white boxes in the same panel show the areas for Figure \ref{8um}.
In the panel (d), the contours for the intensity ratio start at 0.05 with intervals of 0.05.
In the panels (e) and (f), the contours show the total intensities of the [CI] emission in the top left panel, starting at 10 K km s$^{-1}$ with intervals of 10 K km s$^{-1}$.
}
\label{CIMOM0}
\end{figure}

\begin{figure}
\epsscale{0.7}
\plotone{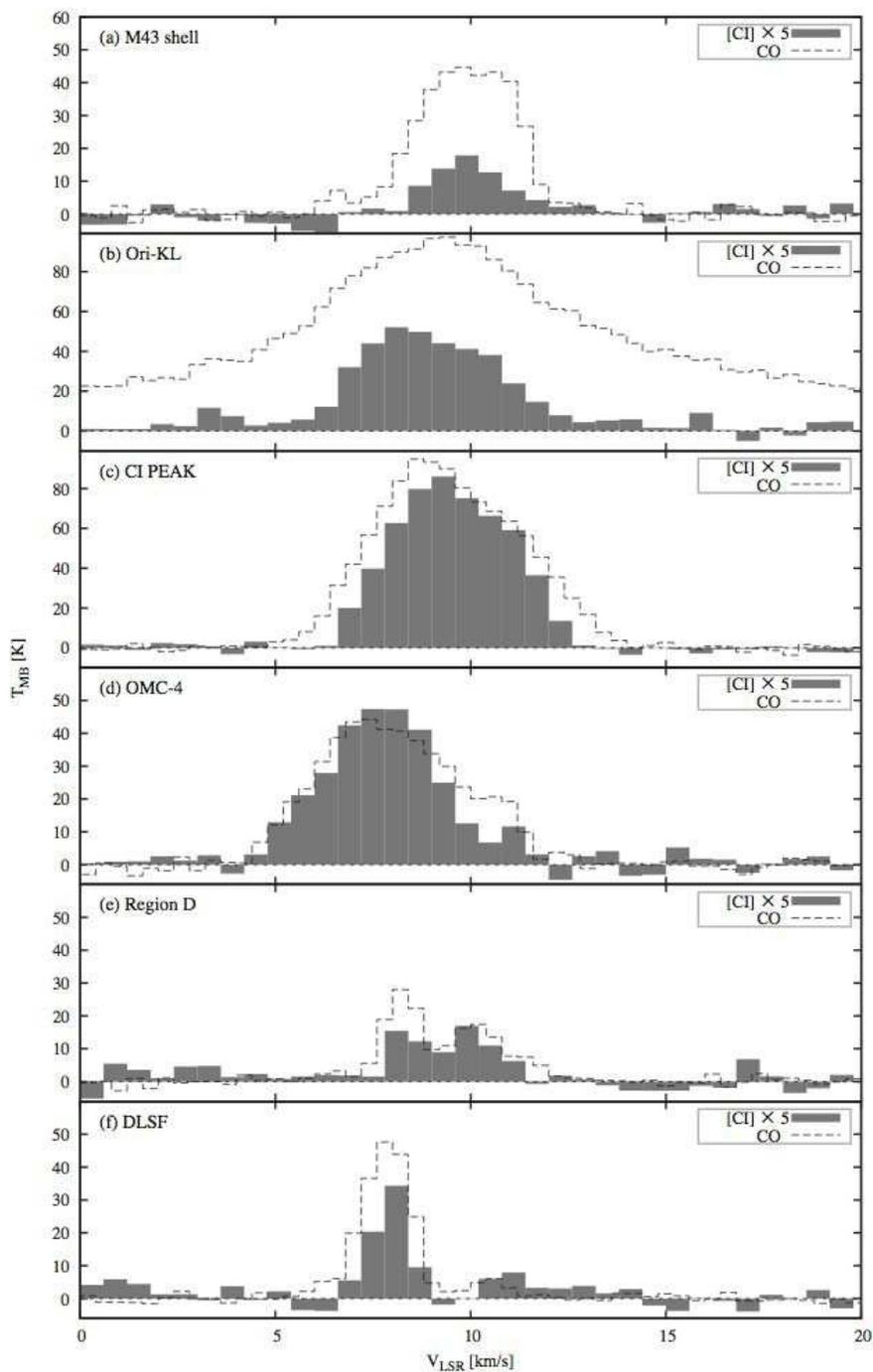}
\caption{Spectra of the [CI] (filled histograms) and $^{12}$CO ($J$=1--0) (dashed lines) emissions at M 43 Shell, Ori-KL, CI PEAK, OMC-4, Region D, and DLSF as shown in Figure \ref{CIMOM0}. 
The velocity resolutions of the [CI] and $^{12}$CO emissions are 0.6 and 0.5 km s$^{-1}$, respectively. 
The rms noise levels of the [CI] and $^{12}$CO emissions are 0.2 K and 0.7 K in $T_{\rm MB}$, respectively. }
\label{spectra}
\end{figure}

\begin{figure}
\begin{center}
\includegraphics[angle=0,scale=0.7]{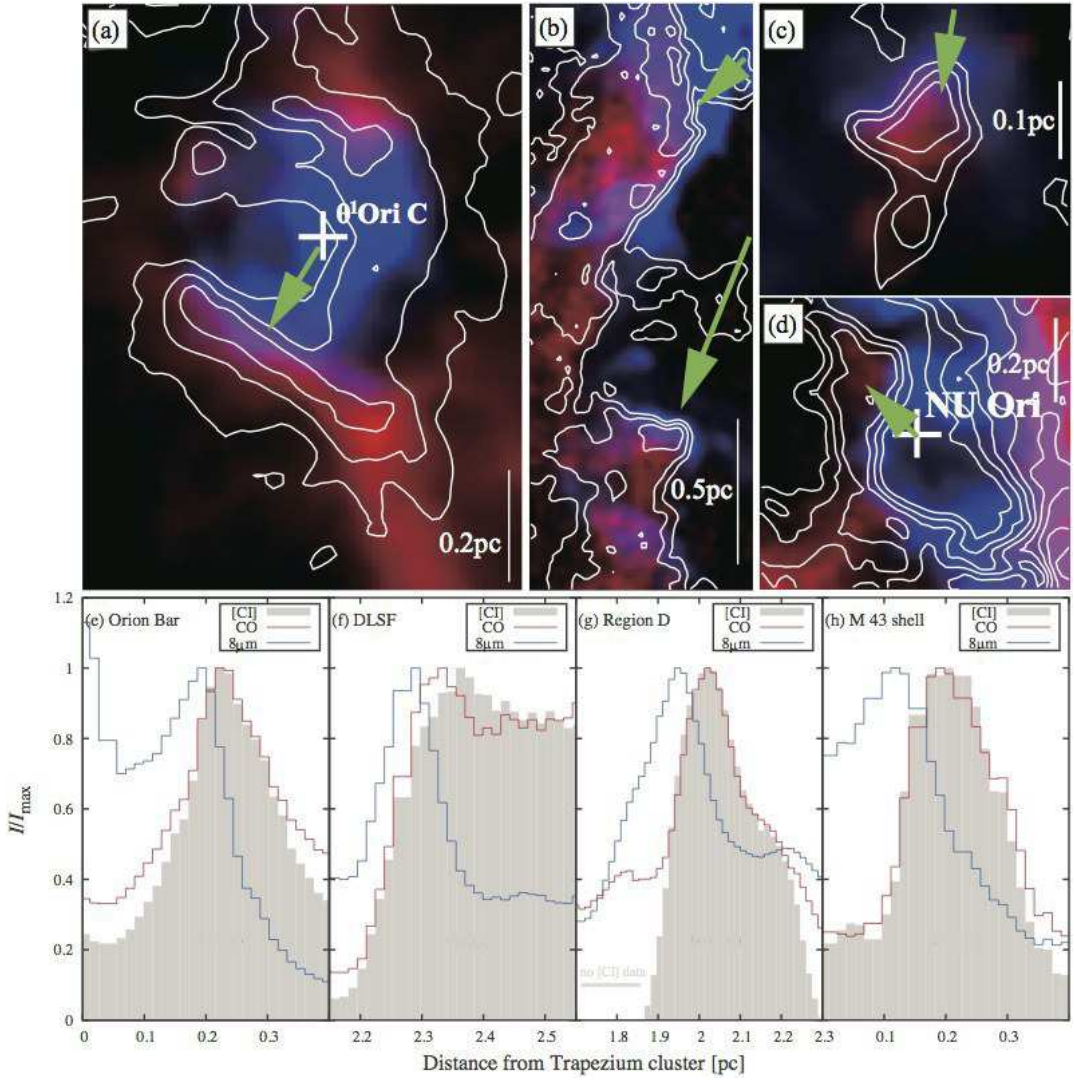}
\end{center}
\caption{
Comparison among the [CI] total intensity (contour), $^{12}$CO peak intensity (red), and MSX 8 $\mu$m intensity (blue) maps in the four PDRs at (a) the Orion Bar, (b) DLSF, (c) Region D , and (d) M 43 Shell.
The bottom four panels show the total intensity profiles of [CI] (grey), $^{12}$CO (red), and 8 $\mu$m (blue) emissions, as a function of the distance from the Trapezium cluster ($\theta^1$OriC) for (e) the Orion Bar, (f) DLSF, and (g) Region D 
and from NU Ori for (h) the M 43 Shell. 
Each intensity profiles was calculated along a line which goes through the position of each exciting star (NU Ori and $\theta^1$OriC)
 and the position in Table \ref{properties}.
The green arrows show directions of the FUV radiation from the exciting stars.
The integration ranges of the [CI] emission in the panels (a)--(d) are 9.3 -- 11.7, 6.9 -- 8.7, 8.1 -- 8.7, and 8.7 -- 11.1 km s$^{-1}$ in $V_{\rm LSR}$, respectively.
In the panel (a), the contours for the $^{12}$CO peak intensity start at 60 K in $T_{\rm MB}$ with intervals of 20 K. In the panels (b) and (d), the contours start at 20 K with intervals of 10 K. In the panel (c), the contours start at 20 K with intervals of 5 K.
}
\label{8um}
\end{figure}

\begin{figure}
\epsscale{1.0}
\plotone{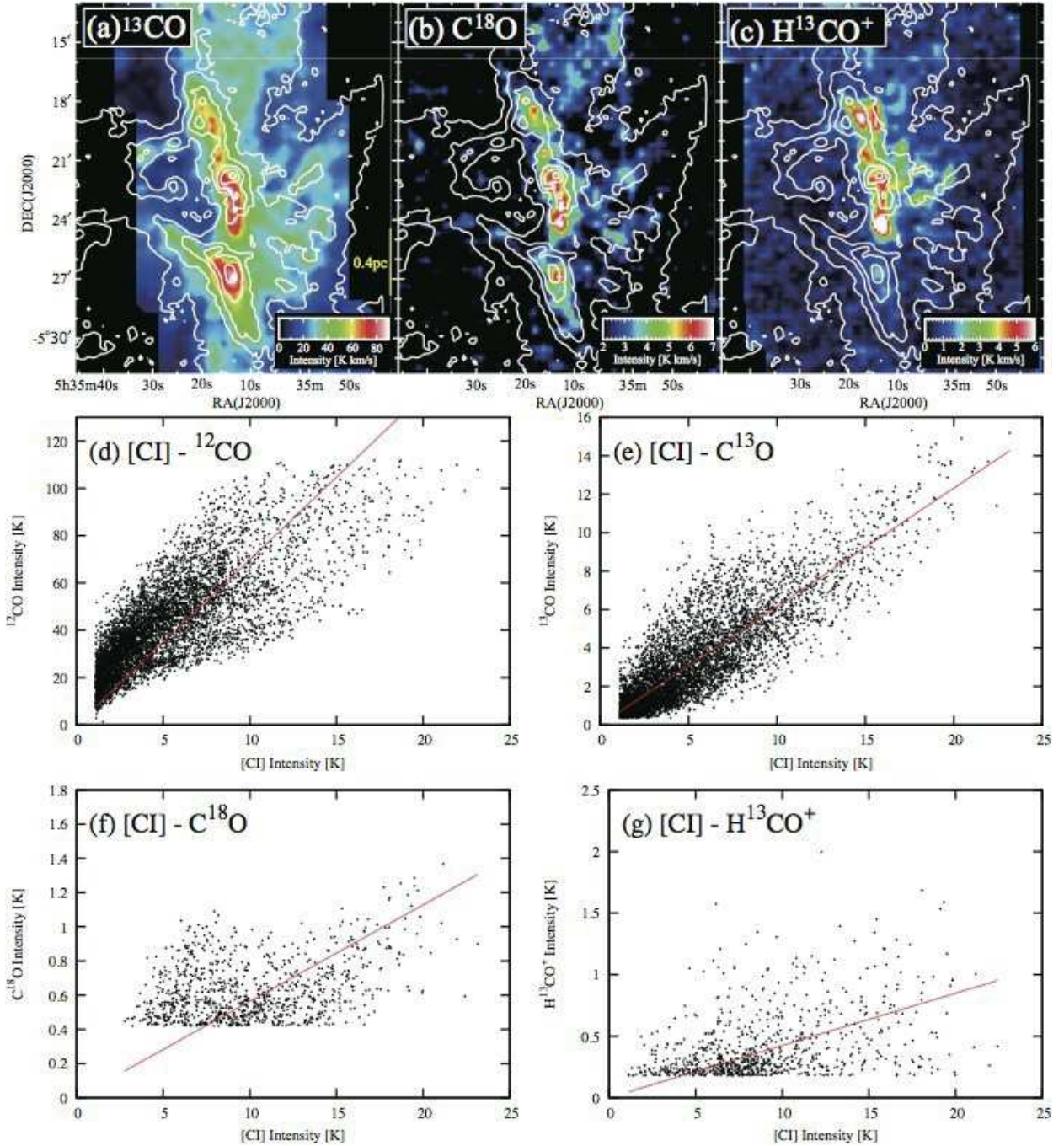}
\caption{Comparison of the [CI] total intensity map in contour with those of the (a) $^{13}$CO ($J$=1--0), (b) C$^{18}$O ($J$=1--0), and (c) H$^{13}$CO$^+$ ($J$=1--0) emissions in color, and point-by-point correlations between the [CI] and (d) $^{12}$CO, (e) $^{13}$CO, (f) C$^{18}$O, and (g) H$^{13}$CO$^+$ emissions. In the panels (a)--(c), the intensity units are K km s$^{-1}$, and all the integration ranges are 3.9 -- 15.9 km s$^{-1}$ in $V_{\rm LSR}$. The contours of the [CI] emission start at 10 K km s$^{-1}$ with intervals of 10 K km s$^{-1}$.  In the panels (d)--(g), only pixels having intensities above the 3 $\sigma$ noise levels in the data cube are plotted. The 1 $\sigma$ noise levels of the [CI], $^{12}$CO, $^{13}$CO, C$^{18}$O, and H$^{13}$CO$^+$ maps are 0.15, 0.4, 0.12, 0.14, and 0.06 K, respectively. The red lines show the results of the least square fitting: $I_{\rm ^{12}CO}$=6.98$\times$$I_{\rm CI}$, $I_{\rm ^{13}CO}$=0.62$\times$$I_{\rm CI}$, $I_{\rm C^{18}O}$=0.06$\times$$I_{\rm CI}$, $I_{\rm H^{13}CO^+}$=0.04$\times$$I_{\rm CI}$.
}
\label{maps}
\end{figure}

Furthermore, the clumpy PDR model is supported by the similarity between the distribution of the [CI] emission and those of the $^{13}$CO ($J$=1--0), C$^{18}$O ($J$=1--0), and H$^{13}$CO$^{+}$ ($J$=1--0) emissions. 
Figure \ref{maps} shows the comparison of the [CI] map with the $^{13}$CO, C$^{18}$O, and H$^{13}$CO$^{+}$ maps and point-by-point correlations between the [CI] and other maps in the OMC-1 region; the $^{13}$CO, C$^{18}$O, and H$^{13}$CO$^{+}$ data are from \citet{Tatematsu93}, \citet{Ikeda07}, and \citet{Ikeda09}, respectively.
The distribution of the [CI] emission is quite similar to that of the $^{13}$CO emission, and is also similar to those of the C$^{18}$O and H$^{13}$CO$^+$ emissions, as suggested by the comparison between the [CI] and $^{12}$CO emissions in Figure \ref{CIMOM0}. 
In the correlation diagrams,  the [CI] emission is the most highly correlated with the $^{13}$CO emission. 
According to \citet{Kawamura98}, we calculated the optical depth of the $^{13}$CO emission using the equation
$\tau({\rm ^{13}CO}) = - {\rm ln} (1-T_{\rm MB}/5.29(J[T_{\rm ex}(K)]-0.164))$, 
where $J[T_{\rm ex}]$ = 1 / exp[5.29/$T_{\rm ex}$] -1. $T_{\rm ex}$ and $T_{\rm MB}$ are the excitation temperature and peak antenna temperature in K. The excitation temperature was assumed to be equal to the $^{12}$CO ($J$=1--0) peak intensity on the assumption that the $^{12}$CO emission is likely to be optically thick. 
The mean optical depth of $^{13}$CO in the map is estimated to be 0.14 $\pm$ 0.06. 
The optical depth is expected to become higher, if there are large temperature gradients in the line of sight \citep{Tauber90}.
Thus, the derived optical depth might be underestimated. 
On the assumption of $T_{\rm ex}$= 20 K, the mean optical depth of $^{13}$CO in the map is, however, estimated to be 0.47 $\pm$ 0.35, suggesting that the $^{13}$CO emission is optically thin.
These results also suggest  that the [CI] emission traces the inner parts of the cloud and supports the clumpy model.

\acknowledgments
We acknowledge the anonymous referee for providing helpful suggestions to improve the paper. 
We also acknowledge the ASTE staffs for both operating ASTE and helping us with the data reduction. 
Observations with ASTE were (in part) carried out remotely from Japan by using NTT's GEMnet2 and its partner R\&E (Research and Education) networks, which are based on AccessNova collaboration of University of Chile, NTT Laboratories, and National Astronomical Observatory of Japan. 
This work was supported by JSPS KAKENHI Grant Number 90610551.
Part of this work was supported by the ANR-11-BS56-010 project gSTARFICHh.

\begin{table}
\begin{center}
\footnotesize
\caption{Physical Properties of the [CI] emission \label{properties}}
\begin{tabular}{lccccccccc}
\tableline\tableline
    Area           & RA & Dec                   & $T_{\rm peak}$ & $V_{\rm peak} $ &  $\int  T_{\rm MB} dV$\tablenotemark{\dagger}  & $T_{\rm ex}$\tablenotemark{\ddagger} & $\tau_{\rm CI}$ & $N_{\rm CI}$ & $I_{\rm CI}/I_{\rm CO}$ \\
 \tableline
              & (J2000) & (J2000)  & (K)                    &   (km s$^{-1}$)        &  (K km s$^{-1}$)   & (K)  &  & ($\times$10$^{17}$ cm$^{-2}$) & \\
\tableline
    M 43 Shell          & 5:35:37.5 &  -5:15:38.3     &    3.6                 &      9.9                   &               8.0              &44.7     &  0.1 & 2.3 & 0.06 \\
    Ori-KL                 & 5:35:14.2 & -5:22:21.5 & 10.4                      & 8.1                       &       49.7                & 97.3      & 0.1 &11.9  & 0.06 \\
    CI PEAK  & 5:35:13.9 & -5:26:45.8    &    17.2               &  9.3                       &       64.0                    &94.7      &  0.2 & 16.1  & 0.15\\
    OMC-4      & 5:35:05.4 & -5:33:45.8       &    9.5                  &    7.5                    &        36.6                  &44.2      & 0.3   &  11.6 & 0.18  \\
    Region D & 5:35:30.5 & -5:40:06.5        &    3.1             &    8.1                  &     9.3              &28.0    &   0.2 & 3.5 & 0.13 \\
    DLSF         & 5:35:51.6 & -5:41:03.3       &    6.9              & 8.1                    &            10.5                  &47.6     & 0.2    &  3.1  & 0.12 \\
\tableline
\tablenotetext{\dagger}{Total integrated intensity over the range 3.75 -- 16.25 km s$^{-1}$ in $V_{\rm LSR}$ at each position}
\tablenotetext{\ddagger}{We consider the $^{12}$CO ($J$=1--0) peak intensity in $T_{\rm MB}$ as the [CI] excitation temperature with the assumption that the $^{12}$CO emission is optically thick.}
\end{tabular}
\end{center}
\end{table}

\end{document}